%% file: ms.tex
\documentclass[useAMS,usenatbib]{mn2e}
\usepackage{epsfig}
\input{macro1}

\title[Luminosity Dependence of Clustering and Velocity Statistics ]
{Luminosity dependence of the spatial and velocity distributions of
galaxies: Semi-analytic models versus the Sloan Digital Sky Survey }

\author[Li, Jing, Kauffmann, B\"orner, Kang \& Wang]{
Cheng Li${^{1,2,3,4}}$\thanks{E-mail: leech@shao.edu.cn},
Y.P. Jing${^{1,4}}$, Guinevere Kauffmann${^{2}}$, Gerhard B\"orner${^{2}}$,
\newauthor Xi Kang${^{1,5}}$, Lan Wang${^{6,2}}$ \\
${^1}$The Partner Group of MPI f\"ur Astrophysik,
      Shanghai Astronomical Observatory,
      Nandan Road 80, Shanghai 200030, China \\
${^2}$Max-Planck-Institut f\"ur Astrophysik,
      Karl-Schwarzschild-Strasse 1, 85748 Garching, Germany \\
${^3}$Center for Astrophysics, University of Science
      and Technology of China, Hefei, Anhui 230026, China \\
${^4}$Joint Institute for Galaxy and Cosmology (JOINGC) of SHAO and USTC\\
${^5}$Max-Planck-Institute for Astronomy, 
      K\"onigstuhl 17, D-69117 Heidelberg, Germany \\
${^6}$Department of Astronomy, Peking University, Beijing 100871, China
}

\begin{document}

\date{ Accepted  ........  Received .......; in  original form ......}

\pagerange{\pageref{firstpage}--\pageref{lastpage}} \pubyear{2006}
\maketitle
\label{firstpage}

\begin{abstract}
By comparing semi-analytic galaxy  catalogues with data from the Sloan
Digital  Sky Survey  (SDSS),  we show  that  current galaxy  formation
models reproduce qualitatively the dependence of galaxy clustering and
pairwise   peculiar  velocities   on  luminosity,   but   some  subtle
discrepancies with the data still remain.  The comparisons are carried
out by constructing  a large set of mock  galaxy redshift surveys that
have the same selection function  as the SDSS Data Release Four (DR4).
The mock  surveys are  based on two  sets of  semi-analytic catalogues
presented by Croton  \etal and Kang \etal.  From  the mock catalogues,
we measure the redshift-space projected two-point correlation function
$w_p(r_p)$,  the power  spectrum $P(k)$  , and  the  pairwise velocity
dispersion   (PVD)   in   Fourier   space  $\sigma_{12}(k)$   and   in
configuration  space  $\sigma_{12}(r_p)$,  for galaxies  in  different
luminosity  intervals. We then  compare these  theoretical predictions
with the measurements derived from  the SDSS DR4.  On large scales and
for  galaxies brighter  than $L^\ast$,  both sets  of  mock catalogues
agree well with the data.   For fainter galaxies, however, both models
predict  stronger  clustering  and  higher  pairwise  velocities  than
observed.  We  demonstrate that  this problem can  be resolved  if the
fraction of faint  satellite galaxies in massive haloes  is reduced by
$\sim30\%$ compared to the model  predictions.  A direct look into the
model galaxy catalogues reveals that a significant fraction ($15\%$) of
faint   galaxies    ($-18<\absmag<-17$)   reside   in    haloes   with
$M_{vir}>10^{13}\msun$,  and this population  is predominantly  red in
colour.  These  faint red  galaxies are responsible  for the  high PVD
values of low-luminosity galaxies on small scales.
\end{abstract}

\begin{keywords}
galaxies: clustering - galaxies: distances and redshifts - large-scale
structure of Universe - cosmology: theory - dark matter
\end{keywords}

\section{Introduction}
The spatial and velocity distributions of galaxies have long served as
important probes of the cosmic density field. Studies of the two-point
correlation  function  (2PCF)  and  the pairwise  velocity  dispersion
(PVD),  reveal  how  galaxies  are  related  to  the  underlying  mass
distribution, thus  providing strong  tests for theoretical  models of
structure and galaxy formation \citep[e.g.][]{Peebles-80,Davis-85}.

Both the  2PCF and  the PVD  can be derived  from redshift  surveys of
galaxies. The studies based on early surveys have established that the
correlation function of L$^\ast$ galaxies is close to a power law over
nearly     four      orders     of     magnitude      in     amplitude
\citep[e.g.][]{Peebles-80}.  It  has also been known  for decades that
the   measured  correlation  of   galaxies  changes   with  luminosity
\citep{Xia-87,   Boerner-91,   Loveday-95}   and  morphological   type
\citep[e.g.  ][]{Davis-Geller-76}.   By taking advantage  of the large
redshift  surveys  assembled  in   recent  years,  in  particular  the
two-degree Field  Galaxy Redshift Survey \citep[2dFGRS;][]{Colless-01}
and  the  Sloan  Digital  Sky  Survey  \citep[SDSS;][]{York-00},  many
authors have studied the dependence  of clustering on a variety galaxy
properties  \citep{Norberg-01,   Norberg-02,  Zehavi-02,  Budavari-03,
Goto-03,  Madgwick-03,  Zehavi-05,  Li-06-wrp}.   These  studies  have
revealed that galaxies  with red colours, bulge-dominated morphologies
and  spectral  types  indicative  of old  stellar  populations  reside
preferentially  in  dense  regions.  Furthermore,  luminous  (massive)
galaxies  cluster  more strongly  than  less  luminous (less  massive)
galaxies,  with   the  luminosity  (mass)   dependence  becoming  more
pronounced  for galaxies  brighter than  $L^\ast$  (the characteristic
luminosity of the \cite{Schechter-76} function).

Measurements of  the PVD have also  been carried out  by many authors,
either   by   modelling   the   redshift   distortion   of   the   2PCF
\citep{Davis-Peebles-83,   Mo-Jing-Borner-93,   Fisher-94,   Zurek-94,
Marzke-95,   Somerville-Davis-Primack-97},   or   by   measuring   the
redshift-space  power spectrum \citep{Jing-Borner-01-obs}.   The early
results  often  varied  significantly   from  one  survey  to  another
\citep{Mo-Jing-Borner-93}.  The PVD of  galaxies in the local Universe
was not  well established until the  work of \cite{Jing-Mo-Boerner-98}
on  the Las  Companas Redshift  Survey.  These results  have now  been
confirmed  by  \citep{Zehavi-02,Zehavi-05}   using  the  SDSS  and  by
\cite{Hawkins-03} using  the 2dFGRS.  \cite{Jing-Borner-04} (hereafter
JB04) presented  the first  determination of the  PVD for  galaxies in
different luminosity  intervals.  This  analysis led to  the discovery
that the PVD exhibits a non-monotonic dependence on galaxy luminosity,
in  that  the  value  of  $\sigma_{12}$ measured  at  $k=1\mpci$  {\em
decreases} as a function of increasing luminosity for galaxies fainter
than $L^\ast$, but  increases again for the most  luminous galaxies in
the sample.  Since  the PVD is an indicator of the  depth of the local
gravitational  potential, this  discovery implies  that  a significant
fraction of faint  galaxies are located in massive  dark matter haloes
that  host  galaxy groups  and  clusters,  but  that the  majority  of
$L^\ast$ galaxies are located in galactic scale haloes.  These results
were recently confirmed by \cite{Li-06-pvd} (hereafter Paper II) using
the second data  release (DR2) of the SDSS.   These authors considered
their results  in conjunction with  the observed luminosity  and color
dependences  of the  two-point correlation  function \citep{Li-06-wrp}
(hereafter Paper I) and concluded that the faint red galaxy population
located in rich clusters was likely to be responsible for the high PVD
values for low-luminosity galaxies on small scales.

A quantitative  understanding of the luminosity dependence  of the PVD
requires a model linking the properties of galaxies to the dark matter
haloes in which  they are found.  One approach is  to carry out N-body
plus                     hydrodynamical                    simulations
\citep[e.g.][]{Katz-Gunn-91,Cen-Ostriker-93,Bryan-94,Navarro-White-94,
Couchman-Thomas-Pearce-95,
Thoul-Weinberg-95,Abel-97,Weinberg-98,Yoshikawa-Jing-Suto-00,Springel-01}.
By numerically solving the gravitational and hydrodynamical equations,
galaxy  formation in  an  expanding  universe can  be  simulated in  a
straightforward way.   However, the  limited dynamic range  in current
hydro/$N$-body simulations and  the limited understanding of important
physical processes such as  star formation and supernova feedback mean
that the  hydrodynamical simulations do  not in general  reproduce the
observed galaxy luminosity  function \citep[e.g.][]{Nagamine-04}. As a
result, these simulations cannot be used to interpret the PVD.

Another  method is  the Halo  Occupation Distribution  (HOD) approach,
which aims  to provide  a statistical description  of how  dark matter
haloes           are          populated           by          galaxies
\citep[e.g.][]{Jing-Mo-Boerner-98,Peacock-Smith-00,Seljak-00,Sheth-01,
Berlind-Weinberg-02,Kang-02,Cooray-Sheth-02}. In  a typical HOD model,
the link between galaxies and dark matter haloes is expressed in terms
of the halo occupation  function $P(N|M)$, which gives the probability
that a  halo of mass $M$  contains $N$ galaxies in  a given luminosity
range.   In   addition,  the  HOD  model  must   specify  the  spatial
distribution of the galaxies  within individual haloes. An alternative
way of describing this link  is in terms of the conditional luminosity
function   $\Phi(L|M)$  \citep[CLF,][]{Yang-Mo-vandenBosch-03},  which
characterises the luminosity distribution of galaxies that reside in a
halo of  mass $M$.  The  HOD approach has  been used to  interpret the
observed dependence  of clustering  on properties such  as luminosity,
colour,          morphology         and          spectral         type
\citep[e.g.][]{Yang-Mo-vandenBosch-03,Yang-04,vandenBosch-Yang-Mo-03,
Yan-Madgwick-White-03,Yan-White-Coil-04,Zehavi-05,Cooray-06}.

JB04 used the HOD models of \cite{Yang-Mo-vandenBosch-03} to construct
mock galaxy catalogues from  N-body simulations. These catalogues were
used to compare  the predicted PVD with the  observations.  They found
that while  the model  provided a successful  match to  the luminosity
function as  well as  the luminosity dependence  of the  clustering on
large scales, it was  unable to reproduce the non-monotonic luminosity
dependence of the PVD  (see Figs. 8 and 9 of JB04  and Fig. 7 of Paper
II). Recently, \citet{Slosar-Seljak-Tasitsiomi-06}  used their own HOD
models to show that the  non-monotonic behaviour can be recovered if a
sufficient number of the faint galaxies are satellite galaxies in high
mass   haloes.   More   recent   studies   by   \cite{Tinker-06}   and
\cite{vandenBosch-06}  also support  this  interpretation.  All  these
studies indicate that the luminosity dependence of the PVD can provide
a strong constraint on theories of galaxy formation.

A third method is to construct semi-analytical models (SAMs) of galaxy
formation
\citep[e.g.][]{White-Frenk-91,Lacey-Silk-91,Kauffmann-White-Guiderdoni-93,
Kauffmann-Nusser-Steinmetz-97,Kauffmann-99,
Cole-94,Cole-00,Somerville-Primack-99}.    This   method  incorporates
parametrised models  to describe the physical  processes that regulate
how stars  form in galaxies as  a function of cosmic  time.  The model
parameters are chosen to  reproduce key observational quantities, such
as  the luminosity  functions of  galaxies in  various  wavebands, the
colour-magnitude   relation   for   early-type   galaxies,   and   the
Tully-Fisher  relation  for  spiral galaxies.   Semi-analytic  models
represent a  powerful way of {\em predicting}  the observed properties
of galaxies.

Two recent SAMs have  been presented by \cite{Kang-05} (hereafter K05)
and  \cite{Croton-06}  (hereafter  C06).   Both models  are  based  on
high-resolution $N$-body simulations  and successfully match a variety
of  observational results.  The  model galaxy  catalogues provided  by
these authors contain information  not only about galaxy distributions
in phase space,  but also about the observed  properties of individual
galaxies  (e.g.   the  absolute  magnitudes in  the  five  photometric
pass-bands of  the SDSS).  In  this paper, we use  these semi-analytic
catalogues to study whether the  luminosity dependence of the 2PCF and
the PVD of  galaxies in the local Universe can  be reproduced in these
models.

As discussed in \cite{Jing-Mo-Boerner-98}, a large set of mock samples
is  essential for this  comparison. The  mock samples  can be  used to
quantify the errors resulting from ``cosmic variance'' effects and and
from systematics in the  estimation methods (e.g. the uncertainties in
the  distribution function  of  peculiar velocities  and  in the  mean
infall velocities  of galaxy pairs).  In this paper, we  construct our
mock galaxy  catalogues that  have the same  selection effects  as the
SDSS Data  Release Four \citep[DR4,][]{Adelman-McCarthy-06}.   To take
into account the  effect of the cosmic variance,  we construct 10 mock
catalogues for each SAM in which the observer is placed at different ,
randomly chosen positions within the simulation box.  Using these mock
catalogues and the same estimation methods used in Papers I and II, we
measure the  redshift-space projected 2PCF  $w_p(r_p)$, the real-space
power  spectrum $P(k)$ and  the PVD  $\sigma_{12}(k)$ for  galaxies in
different luminosity intervals.

Observational  results from the  SDSS have  already been  presented in
Papers I and II, using a  sample of $\sim$ 200,000 galaxies drawn from
the  SDSS  Data  Release  Two  (DR2).   Here  we  re-compute  all  the
clustering statistics using the SDSS DR4 in order to take advantage of
the larger  number of galaxies in  the newer release.  We also measure
$\sigma_{12}(r_p)$, the  PVD in configuration space, in  order to make
comparisons        with       recent        HOD        models       of
\citet{Slosar-Seljak-Tasitsiomi-06} and \cite{Tinker-06}.

In   the   following    sections   we   describe   the   observational
measurements(\S2),    the     procedure    for    constructing    mock
catalogues(\S3), the comparison  between models and observations(\S4),
the mock  experiments to bring  the models into better  agreement with
the  data(\S5), and  the nature  of the  luminosity dependence  of the
PVD(\S6).  In \S7, we  summarise our results, discuss the implications
for  the models,  and suggest  possible improvements  both  for future
observations and models.

Throughout this paper, we assume a cosmological model with the density
parameter  $\Omega_0=0.3$ and  cosmological  constant $\Lambda_0=0.7$.
In  the  C06  SAM  model,  the cosmological  parameters  are  slightly
different from  these adopted values.  To properly compare  this model
with  the observations,  we  calculate the  positions, redshifts,  and
apparent  magnitudes  for mock  galaxies  using  the C06  cosmological
parameters. In the analysis of mock galaxy clustering, we use the same
cosmological  parameters  as  in  the analysis  of  the  observational
clustering.   A    Hubble   constant   $h=1$,   in    units   of   100
kms$^{-1}$Mpc$^{-1}$, is assumed  throughout this paper when computing
absolute magnitudes.

\section{Observational measurements}
\label{sec:obs}

\subsection{Samples}

\begin{table}
\caption{Luminosity samples selected from the NYU-VAGC {\tt Sample dr4}.}
\label{tbl:samples}
\begin{center}
\begin{tabular}{lccr}\hline\hline
& \multicolumn{2}{c}{$M_{^{0.1}r}$}  & {\sc Number  of} \\ \cline{2-3}
{\sc Sample}&  {\sc Range} & {\sc  Median} & {\sc  Galaxies} \\ \hline
L1...........&$[-18.0,-17.0)$&-17.59             &7090              \\
L2...........&$[-18.5,-17.5)$&-18.09             &11992             \\
L3...........&$[-19.0,-18.0)$&-18.59             &20571             \\
L4...........&$[-19.5,-18.5)$&-19.11             &38203             \\
L5...........&$[-20.0,-19.0)$&-19.58             &66737             \\
L6...........&$[-20.5,-19.5)$&-20.05             &98589             \\
L7...........&$[-21.0,-20.0)$&-20.52             &121822            \\
L8...........&$[-21.5,-20.5)$&-20.95             &113449            \\
L9...........&$[-22.0,-21.0)$&-21.38             &70499             \\
L10......... &$[-22.5,-21.5)$&-21.80             &27427             \\
L11......... &$[-23.0,-22.0)$&-22.22             &6085       \\ \hline
\end{tabular}
\end{center}
\end{table}

The  SDSS is  the  most ambitious  optical  imaging and  spectroscopic
survey to date. The survey goals are to obtain photometry of a quarter
of  the sky and  spectra of  nearly one  million objects.   Imaging is
obtained    in     the    {\em    u,    g,    r,     i,    z}    bands
\citep{Fukugita-96,Smith-02,Ivezic-04}  with a  special  purpose drift
scan camera  \citep{Gunn-98} mounted  on the SDSS  2.5~meter telescope
\citep{Gunn-06}  at Apache  Point Observatory.   The imaging  data are
photometrically    \citep{Hogg-01,Tucker-06}    and    astrometrically
\citep{Pier-03} calibrated,  and used to  select spectroscopic targets
for the main galaxy sample \citep{Strauss-02}, the luminous red galaxy
sample     \citep{Eisenstein-01},     and     the    quasar     sample
\citep{Richards-02}.  Spectroscopic fibres are assigned to the targets
using an efficient tiling  algorithm designed to optimise completeness
\citep{Blanton-03-tiling}.  The details of  the survey strategy can be
found in  \citet{York-00} and  an overview of  the data  pipelines and
products   is    provided   in   the   Early    Data   Release   paper
\citep{Stoughton-02}. More  details on the photometric  pipeline can be
found in \citet{Lupton-01}.

Papers  I and  II  presented the  measurements  of the  redshift-space
projected 2PCF  $w_p(r_p)$, the  real-space power spectrum  $P(k)$ and
the PVD  $\sigma_{12}(k)$ for different classes of  galaxies. In those
papers,  we  used  the  New  York  University  Value  Added  Catalogue
(NYU-VAGC)
\footnote{http://wassup.physics.nyu.edu/vagc/},  which is  a catalogue
of  local   galaxies  (mostly  below   $z\approx0.3$)  constructed  by
\citet{Blanton-05-VAGC}  based on  the SDSS  DR2.  Here we  use a  new
version of  the NYU-VAGC  ({\tt Sample dr4}),  which is based  on SDSS
DR4, to re-determine these statistics, but as a function of luminosity
only. The NYU-VAGC is described in detail in \citet{Blanton-05-VAGC}.

From  {\tt Sample  dr4},  we construct  11  luminosity subsamples,  as
listed  in   Table~\ref{tbl:samples}.  We  select   all  objects  with
$14.5<r<17.6$ that are identified as galaxies in the Main sample (note
that   $r$-band   magnitude    has   been   corrected   for   galactic
extinction).  We also  restrict  the galaxies  to  the redshift  range
$0.01\leq    z\leq0.3$,    and    the   absolute    magnitude    range
$-23<M_{^{0.1}r}<-17$.  Here, $M_{^{0.1}r}$  is the  $r$-band absolute
magnitude corrected to its $z=0.1$ value using the $K$-correction code
of \cite{Blanton-03-Kcorrection} and the luminosity evolution model of
\cite{Blanton-03-LF}.   The  resulting  sample  includes  a  total  of
292,782 galaxies, which are  then divided into subsamples according to
absolute magnitude.  Each subsample  includes galaxies in  an absolute
magnitude   interval  of  1   magnitude,  with   successive  subsamples
overlapping by  0.5 magnitude.  This sample selection  is identical to
that  in Paper  I,  except  that we  have  adopted slightly  different
apparent and absolute magnitudes  limits.  We do not consider galaxies
fainter  than $M_{^{0.1}r}=-17$,  because the  volume covered  by such
faint  samples are very  small and  the results  are subject  to large
errors from cosmic variance (see for  example Fig. 6 of Paper I).  The
faint apparent  magnitude limit of 17.6  is chosen to  yield a uniform
galaxy sample that is complete over the entire area of the survey.

\subsection{Methods}

Our methodology for  computing $w_p(r_p)$, $P(k)$ and $\sigma_{12}(k)$
in  the SDSS  has been  described in  detail in  Papers I  and  II. We
present below  a brief description and  the reader is  referred to the
earlier papers for details.

For  each  subsample, the  redshift-space  2PCF $\xi^{s}(r_p,\pi)$  is
measured using  the \cite{Hamilton-93} estimator.   The redshift-space
projected   2PCF   $w_p(r_p)$  is   then   estimated  by   integrating
$\xi^{(s)}(r_p,\pi)$  along  the  line-of-sight direction  $\pi$  with
$|\pi|$ ranging from 0 to  40 $\mpch$.  Random samples are constructed
in which  the redshift selection function is  explicitly modelled using
the observed  luminosity function.   We have also  corrected carefully
for  the  effect of  fibre  collisions  (see \citep[][hereafter  Paper
III]{Li-06-agn} for a detailed description).

From   $\xi^{(s)}(r_p,\pi)$,  we   obtain  for   each   subsample  the
redshift-space  power spectrum  $P^{(s)}(k,\mu)$.   We then  determine
simultaneously the power spectrum  $P(k)$ and the PVD $\sigma_{12}(k)$
by modelling the measured $P^{(s)}(k,\mu)$ using the relation
\begin{equation}\label{eqn:redpow}
P^{(s)}(k,\mu)    =    P(k)(1+\beta    \mu^2)^2   \frac{1}{1+(k    \mu
\sigma_{12}(k))^2}.
\end{equation}
Here $k$ is the wavenumber, $\mu$  the cosine of the angle between the
wavevector  and the  line of  sight, and  $\beta$ the  linear redshift
distortion parameter.  In the  equation above, the  first term  is the
power  spectrum, the  second  term is  the  Kaiser linear  compression
effect (Kaiser 1987), and the  third term is the damping effect caused
by  the random motion  of the  galaxies. In  the computation,  we have
fixed the  linear redshift  distortion parameter $\beta=0.45$.   As we
have shown  in Paper II, our $\sigma_{12}(k)$  measurements are robust
to reasonable changes of the $\beta$ values.

In   addition,  we   also   compute  the   configuration  space   PVD,
 $\sigma_{12}(r_p)$,   which  is   estimated  by   modelling  redshift
 distortions in  the 2PCF.   This method relies  on the fact  that the
 peculiar motions  of galaxies change  only their radial  distances in
 redshift space.   Thus the information for  peculiar velocities along
 the line  of sight can  be recovered by modelling  the redshift-space
 2PCF  $\xi^{(s)}(r_p,\pi)$ as  a convolution  of the  real-space 2PCF
 $\xi(r)$  with the  distribution  function of  the pairwise  velocity
 $f(v_{12})$:
\begin{equation}\label{eqn:xipv}
\xi^{(s)}(r_p,\pi)= 
\int f(v_{12})\xi\left(\sqrt{r_p^2+(\pi-v_{12})^2}\right)dv_{12},
\end{equation}
where $v_{12}=v_{12}(r_p,\pi)$ is  the pairwise peculiar velocity. The
real-space  correlation   function  $\xi(r)$  is   inferred  from  the
projected  2PCF  $w_p(r_p)$,  which  is  a simple  Abel  transform  of
$\xi(r)$. An exponential form is adopted for $f(v_{12})$:
\begin{equation}\label{eqn:expfv}
f(v_{12})=\frac{1}{\sqrt{2}\sigma_{12}}
\exp\left(-\frac{\sqrt{2}}{\sigma_{12}}
\left|v_{12}-\overline{v_{12}}\right|\right)
\end{equation}
where  $\overline{v_{12}}$  is  the  mean  and  $\sigma_{12}$  is  the
dispersion of the  one-dimensional peculiar velocities.  Assuming the
infall      model       for      $\overline{v_{12}}$      used      by
\cite{Jing-Mo-Boerner-98}, the PVD is  then estimated as a function of
the   projected   separation   $r_p$   by   comparing   the   observed
$\xi^{(s)}(r_p,\pi)$ with the modelled one.

\subsection{Results}

\begin{figure*}
\vspace{-0.5cm}
\centerline{\epsfig{figure=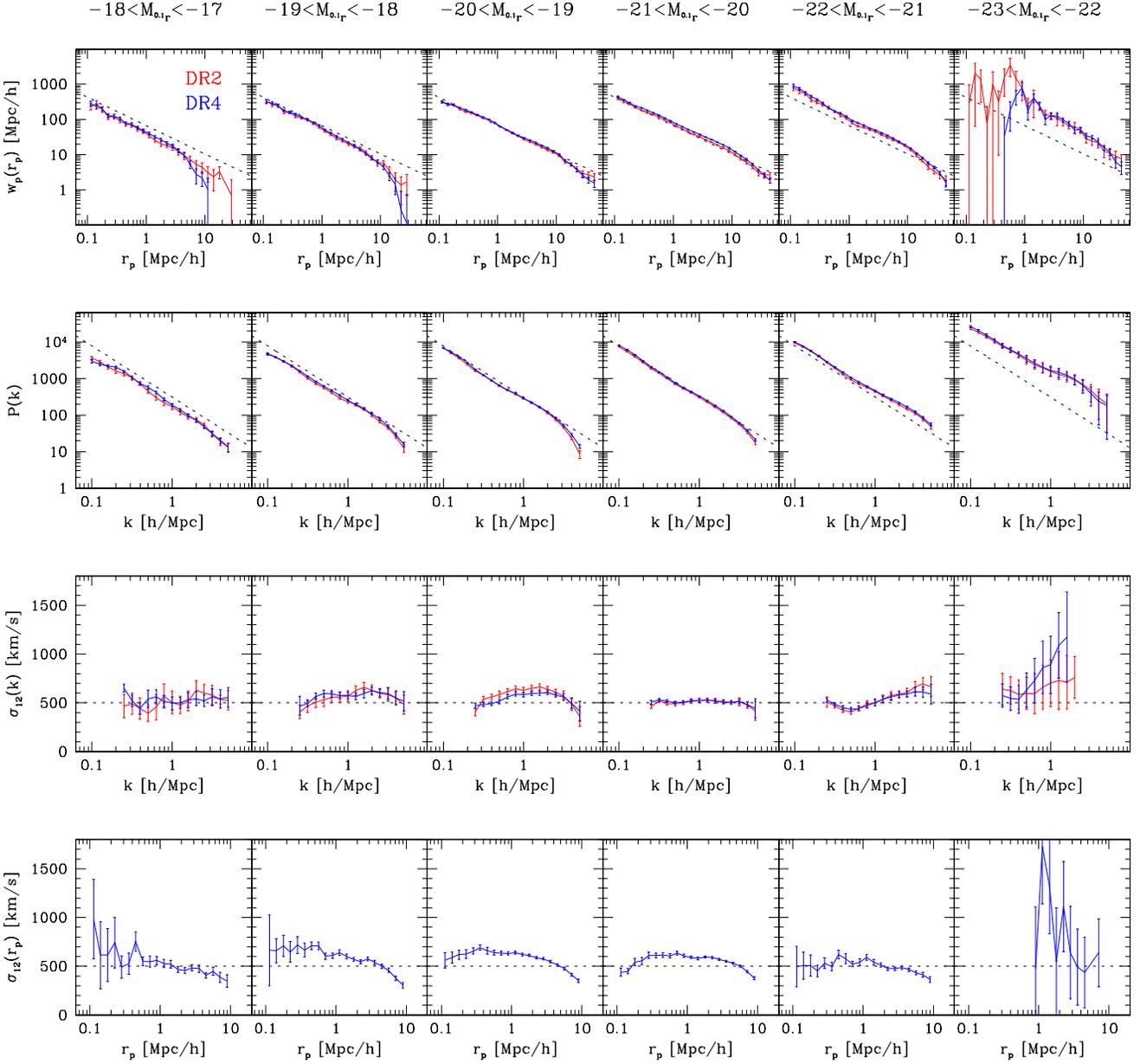,width=\hdsize}}
\vspace{-0.5cm}
\caption{Clustering and velocity  statistics for galaxies with various
luminosities.   Panels  from left  to  right  correspond to  different
luminosity intervals, as  indicated at the top of  the figure.  Panels
from top  to bottom correspond to different  statistics: the projected
2PCF  $w_p(r_p)$,  the  real  space  power spectrum  $P(k)$,  the  PVD
measured   in  Fourier   space  $\sigma_{12}(k)$,   and  the   PVD  in
configuration  space  $\sigma_{12}(r_p)$.   Blue  and  red  lines  are
measured from the SDSS DR4  and the SDSS DR2 respectively.  The dashed
lines are plotted for reference, which  are the same in each row (from
top   to    bottom):   the   line    corresponding   to   $\xi(r)=(r/5
h^{-1}$Mpc$)^{-1.8}$, the power  spectrum $P(k)=(60/k)^{1.4}$, and (in
both the bottom two rows) the line for $\sigma_{12}=500\kms$.}
\label{fig:dr2_dr4}
\end{figure*}
\begin{figure*}
\vspace{-0.2cm}
\centerline{\epsfig{figure=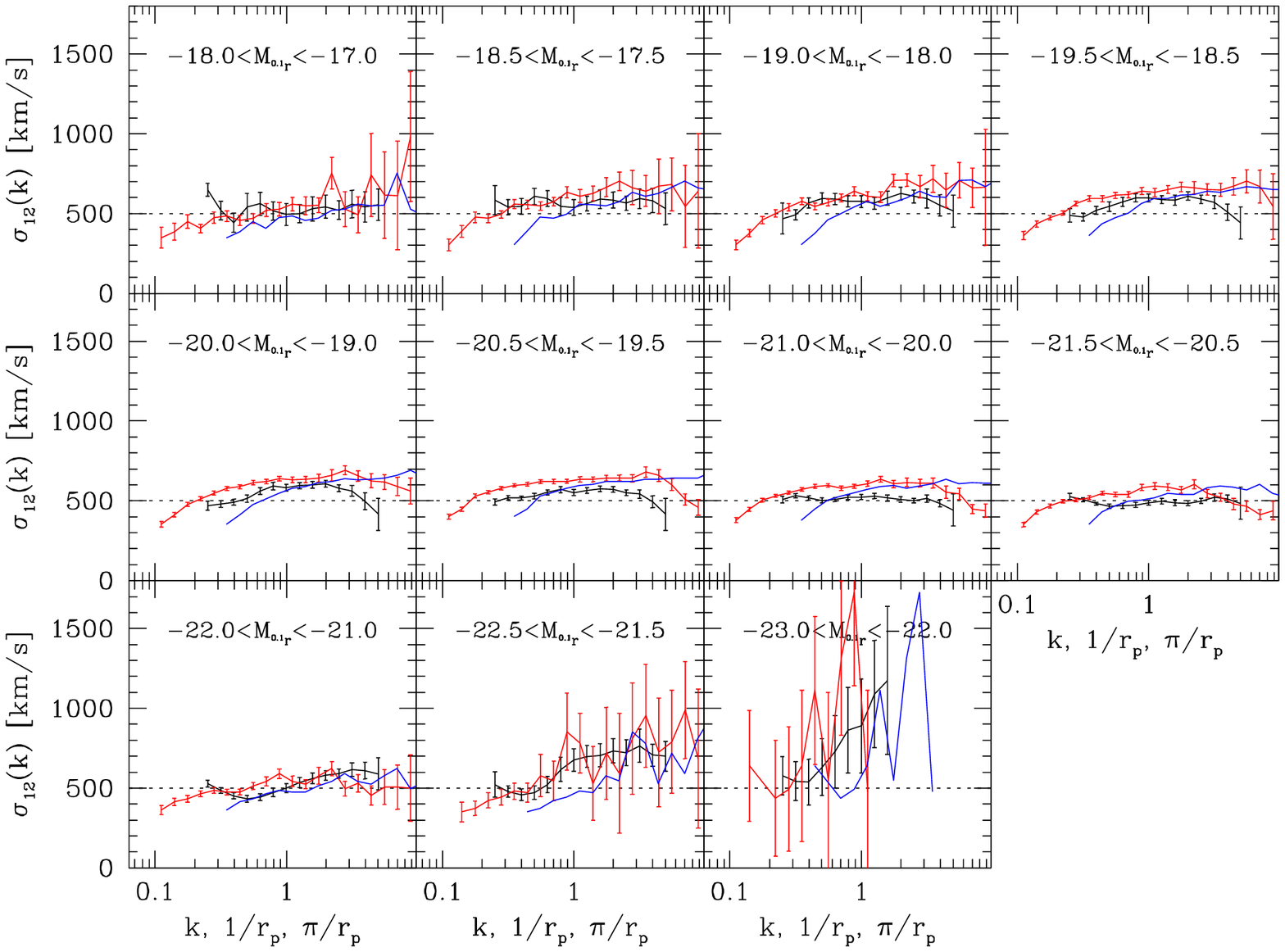,width=\hdsize}}
\vspace{-4.7cm}
\caption{Comparison  of  PVD  as  measured  in Fourier  space  and  in
configuration space for galaxies in different luminosity intervals, as
indicated in each panel. Black  lines plot $\sigma_{12}$ as a function
of $k$. The PVDs measured in configuration space are plotted in red as
a  function   of  $1/r_p$  and  in   blue  as  a   function  of  ${\rm
\pi}/r_p$. The dashed  line in each panel represents  the PVD value of
500$\kms$.}
\label{fig:sig_k}
\end{figure*}

Using the  samples listed  in Table~\ref{tbl:samples} and  the methods
described above,  we have derived  the the projected  2PCF $w_p(r_p)$,
the power spectrum $P(k)$,  the PVD in Fourier space $\sigma_{12}(k)$,
and the  PVD in  configuration space $\sigma_{12}(r_p)$.   The results
are  shown in  Figure~\ref{fig:dr2_dr4}.   Panels from  left to  right
correspond to the  six luminosity subsamples (samples L1,  L3, L5, L7,
L9  and L11  in  Table~\ref{tbl:samples}), while  panels  from top  to
bottom correspond  to the  different clustering statistics.   The blue
and red lines compare the results  obtained from the DR4 and DR2.  The
two data releases agree quite well, except for $\sigma_{12}(k)$ in the
brightest  luminosity sample,  where  the DR4  measurements on  scales
$k>0.5\mpci$ are larger  , but still within the error  bars of the DR2
measurements.

A comparison  of the PVD for  the two different  estimation methods is
shown in  Figure~\ref{fig:sig_k} for  all 11 luminosity  samples.  The
$k$-space  measurements $\sigma_{12}(k)$  are plotted  in  black.  The
PVDs in configuration space $\sigma_{12}(r_p)$ are plotted in red as a
function of $1/r_p$ and in blue  as a function of ${\rm \pi}/r_p$.  We
see  that,  if the  relation  $k=1/r_p$  is  used in  the  comparison,
$\sigma_{12}(r_p)$ and  $\sigma_{12}(k)$ agree well  within error bars
both in  shape and  in amplitude, for  galaxies fainter than  $-19$ or
brighter than $-21$.  For galaxies around $L^\ast$, $\sigma_{12}(r_p)$
is systematically higher  than $\sigma_{12}(k)$, by up to  30 per cent
on intermediate scales. Using ${\rm \pi}/r_p$ for $k$ does not improve
the agreement  between the two  quantities. It is not  surprising that
there are differences between the results, because the PVD $\sigma(r)$
in 3D configuration space is not a constant. Our results indicate that
it is important that the PVDs from the semi-analytic model be computed
in exactly the same manner as is done in the observations.

\section{Mock catalogues}
\label{sec:mock}

\subsection{SAMs and model catalogues}

In  this paper,  we  compare the  clustering  and velocity  statistics
predicted  by  the  semi-analytic  models  with  the  observations  by
constructing a  large set  of mock galaxy  samples that have  the same
selection effects as  the SDSS DR4.  We use  two sets of semi-analytic
catalogues of galaxies at $z=0$, provided by C06 and K05, to construct
our mock catalogues.

The  semi-analytic  catalogues  of  C06  were  constructed  using  the
Millennium  Run \citep{Springel-05},  a very  large simulation  of the
concordance  $\Lambda$CDM  cosmogony  with  $10^{10}$  particles.  The
relevant   cosmological   parameters   are   the   density   parameter
$\Omega_m=0.25$, the  cosmological constant $\Omega_\Lambda=0.75$, and
the  amplitude  of  the  power spectrum  $\sigma_8=0.9$.   The  chosen
simulation volume  is a periodic  box of size $L_{box}=500\mpch$  on a
side, which  implies a particle mass  of $8.6\times10^8\msunh$.  After
finding  haloes and  subhaloes at  all output  snapshots  and building
merging trees that  describe how haloes grow as  the universe evolves,
C06 implemented  a model  to simulate the  formation and  evolution of
galaxies  and their  central  supermassive black  holes.  Their  model
closely  matches many  observations, including  the  galaxy luminosity
function,  galaxy colour distributions,  the Tully-Fisher  relation of
spirals,  the  colour-magnitude  relation  of ellipticals,  the  bulge
mass-black  hole mass  relation, and  the volume-averaged  cosmic star
formation rate.  The  models yield a number of  useful quantities that
can   be   directly    compared   with   observations   at   different
redshifts. These include positions  in phase space, total luminosities
and bulge  luminosities in various  bands, stellar masses ,  cold, hot
and ejected gas  mass, black hole mass, and  star formation rate.  The
semi-analytic  galaxy catalogue  used  here is  publicly available  at
http://www.mpa-garching.mpg.de/galform/agnpaper  and   it  includes  a
total  of $\sim9\times10^{6}$ galaxies  at redshift  zero in  the full
simulation box. The catalogue  is complete down to $M_r-5\log h=-16.6$
and to $M_B-5\log h=-15.6$.

Using  the semi-analytical  approach, K05  also carried  out a  set of
semi-analytic  galaxy catalogues  by  modelling galaxy  formation in  a
series of high-resolution  $N$-body simulations.  The simulations used
in  their study  have been  carried out  with the  vectorised parallel
P$^3$M  code  \citep{Jing-Suto-02},  considering boxes  with  periodic
boundary  conditions  in a  concordance  $\Lambda$CDM cosmology.   The
cosmological  parameters  $\Omega_m=0.3$  and $\Omega_\lambda=0.7$  are
slightly different from  those of C06. There are  $512^3$ particles in
the     simulation     box     of    $L_{box}=100\mpch$     ($L_{100}$
simulation). Although the simulation is much smaller than that of C06,
the  mass resolution is  comparable.  The  galaxy formation  model has
been  updated to  include supermassive  black hole  formation  and AGN
energy  feedback, as described  in \cite{Kang-Jing-Silk-06}.   The SAM
model of  K05 can also  match many observations, e.g.   the luminosity
functions of  galaxies in various wavebands redder  than the $u$-band,
the main features in the  observed colour distribution of galaxies, the
colour-magnitude  relation  for elliptical  galaxies  in clusters,  the
metallicity-luminosity  relation   and  metallicity-rotation  velocity
relation  of spiral  galaxies,  and the  gas  fraction in  present-day
spiral galaxies.

In order to study the clustering  of galaxies on large scales, we will
use  a  simulation  of  $512^3$  particles  and  box  size  $300\mpch$
($L_{300}$ simulation)  with the  same cosmological parameters  as the
smaller box. Because of its poor mass resolution, we do not follow the
formation  histories  of  galaxies  in this  simulation.  Instead,  we
combine the $L_{100}$ simulation and a set of resimulations of massive
clusters  of  $\sim 10^{15}\msun$  (see  K05),  and  use these  higher
resolution  simulations  to populate  the  dark  matter  halos in  the
$L_{300}$  simulation.  In detail,  for  each  halo  in the  $L_{300}$
simulation, we  select an  halo from the  $L_{100}$ simulation  or the
cluster resimulations that  is closest in mass.  The  galaxies of this
matching halo will  be placed into the $L_{300}$  simulation halo. All
physical properties as well as the relative position and velocity with
respect to centre of halo mass are kept the same.

The  Kang  et al.  and  C06  are similar,  but  there  are still  many
differences  in the details  of the  implementation. For  example, C06
allowed starbursts  to be triggered  during minor mergers.  The energy
released by gas accretion onto  the central supermassive black hole is
also slightly different in  the two implementations. The parameters of
the  star  formation laws  and  even  of  the cosmological  models  are
different. These differences make it interesting for us to compare the
clustering of galaxies in the two SAM implementations.

\begin{figure}
\vspace{-0.2cm}
\centerline{\epsfig{figure=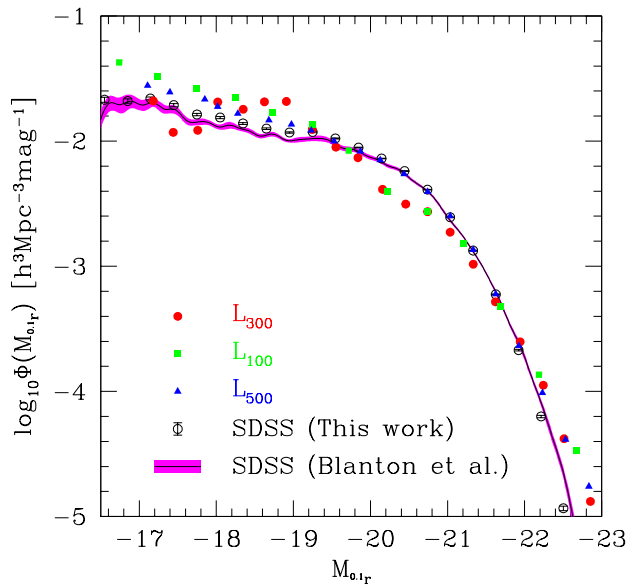,width=\hssize}}
\vspace{-0.2cm}
\caption{Galaxy luminosity  function at $^{0.1}r$-band.  Triangles are
for the  $L_{500}$ SAM catalogue  of Croton et al.(2006).  Squares and
filled  circles  are  respectively  for the  $L_{100}$  and  $L_{300}$
catalogues of Kang  et al. (2005). The line  surrounded by the magenta
band  plots   the  observational   result  presented  by   Blanton  et
al. (2003b) based  on the first data release of  the SDSS; the magenta
band indicates its error.  Open  circles with error bars represent the
result obtained in this paper with the SDSS DR4.}
\label{fig:lf}
\end{figure}
\begin{figure*}
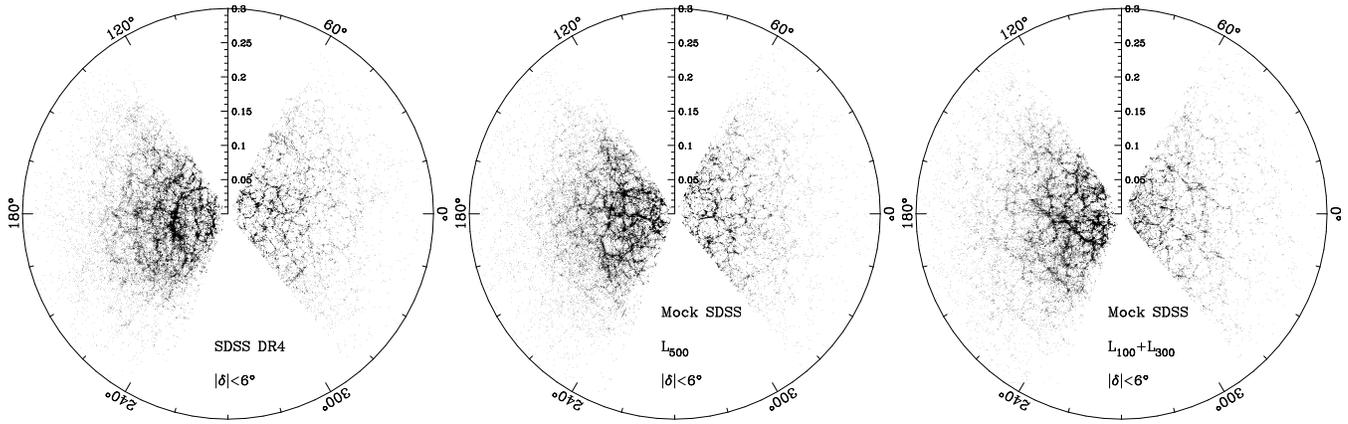

\centerline{\epsfig{figure=f4a.ps,width=0.33\hdsize}
            \epsfig{figure=f4b.ps,width=0.33\hdsize}
            \epsfig{figure=f4c.ps,width=0.33\hdsize}}
\caption{Equatorial distribution  of right ascension  and redshift for
galaxies within $6^\circ$ of the equator in the SDSS (left) and in our
mock catalogues.}
\label{fig:ra_red}
\end{figure*}

Figure~\ref{fig:lf} shows  the $^{0.1}r$-band luminosity  function for
galaxies  in  the semi-analytical  catalogues,  compared  to the  SDSS
observations presented  in \citet{Blanton-03-LF}.  The  SAMs reproduce
the observed luminosity function reasonably well , although they still
predict    too    many    faint   ($M_{^{0.1}r}>-19$)    and    bright
($M_{^{0.1}r}<-22$) galaxies.   It is worth noting  that the $L_{100}$
and  $L_{300}$   catalogues  contain  fewer   $L^\ast$  galaxies  than
observed.

We have also recomputed  the observed galaxy luminosity function using
our SDSS DR4 sample.  We  have corrected for the volume incompleteness
by weighting  each galaxy by  a factor of  $V_{survey}/V_{max}$, where
$V_{survey}$ is the volume of  the sample and $V_{max}$ is the maximum
volume over  which the  galaxy could be  observed within  the redshift
range and the apparent magnitude range of the sample. To determine the
$V_{max}$,    we   have    used   the    {\tt   kcorrect}    code   of
\citet{Blanton-03-Kcorrection} to compute  for each galaxy a $z_{min}$
and a  $z_{max}$, the  redshifts at which  the galaxy would  reach the
bright and  the faint $r$-band  magnitude limits.  Our  measurement is
also shown in  Figure~\ref{fig:lf}, and it agrees quite  well with the
result of  \citet{Blanton-03-LF}. The  errors are estimated  using the
bootstrap resampling technique \citep{Barrow-Bhavsar-Sonoda-84}.

\subsection{Constructing mock galaxy redshift surveys}

\begin{figure*}
\vspace{-0.5cm}
\centerline{\epsfig{figure=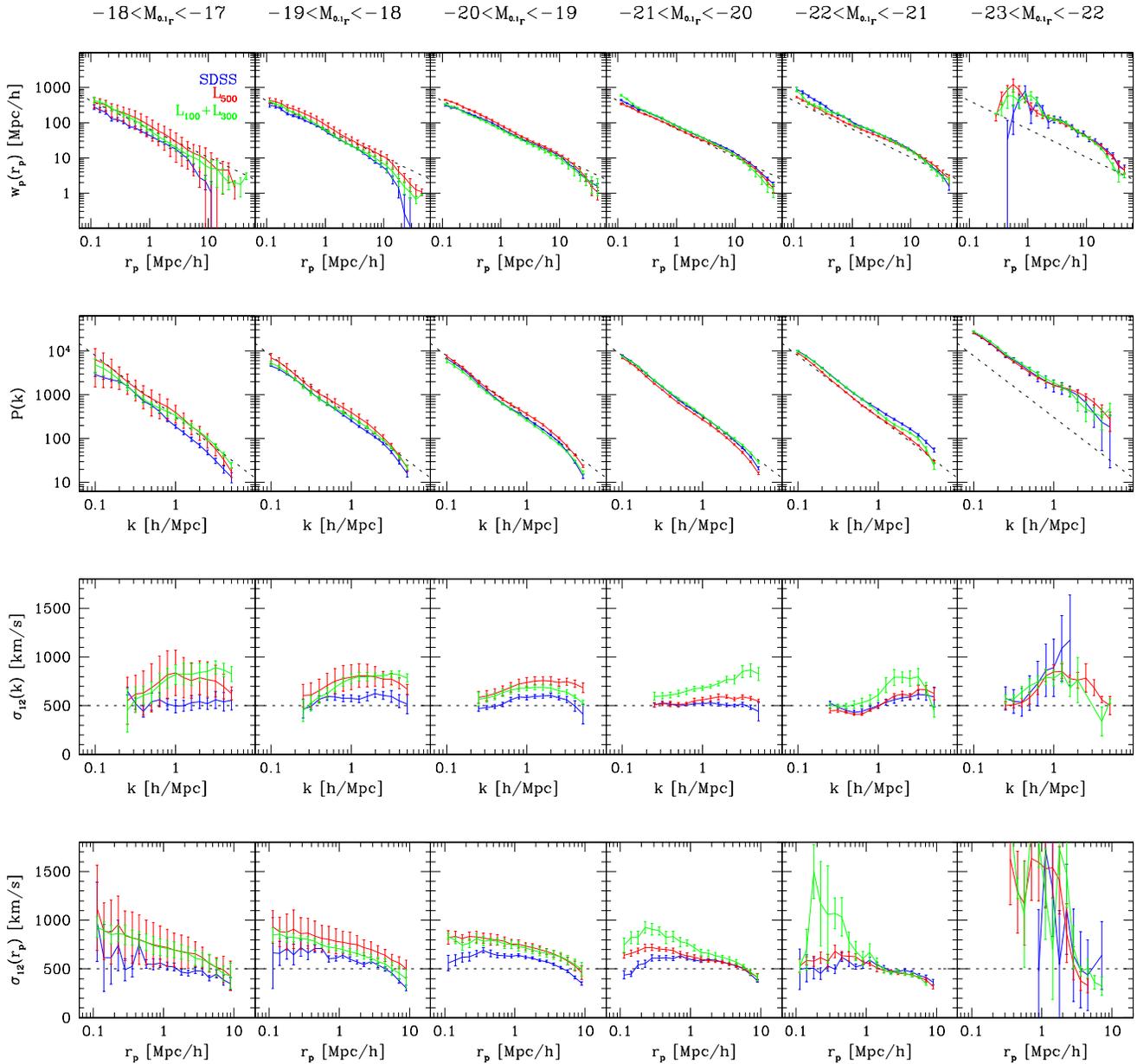,width=\hdsize}}
\vspace{-0.5cm}
\caption{Comparison  of  the  clustering  and velocity  statistics  as
measured  from mock  catalogues and  as  observed from  SDSS DR4,  for
galaxies  with  various  luminosities.   Panels  from  left  to  right
correspond to  different luminosity intervals, as  indicated above the
top  panels.  Panels  from  top  to  bottom  correspond  to  different
statistics:      $w_p(r_p)$,     $P(k)$,      $\sigma_{12}(k)$     and
$\sigma_{12}(r_p)$.   Red  and   green  lines  represent  the  average
measurement from the $L_{500}$  and the $L_{100}+L_{300}$ mock samples
respectively.  The  error bars indicate the uncertainty  due to cosmic
variance as  estimated from 10 different mock  catalogues.  Blue lines
plot the observational  results.  The dashed lines are  the same as in
Figure~\ref{fig:dr2_dr4}.}
\label{fig:dr4_mock}
\end{figure*}

We aim  to construct mock galaxy  redshift surveys that  have the same
observational selection  effects as the SDSS DR4.   A detailed account
of the  observational selection effects accompanies  with the NYU-VAGC
release.   Our methodology  of constructing  mock SDSS  catalogues has
been  described in  detail in  Paper  III. First,  we create  $n\times
n\times  n$ replications  of  the simulation  box  which has  periodic
boundary conditions, and place  a virtual observer randomly within the
central box.   Here $n$ is  chosen so that  the required depth  can be
achieved  in all  directions  for  the observer.   Next,  we define  a
($\alpha$,$\delta$)-coordinate frame and  remove all galaxies that lie
outside  the survey  region.   We  then compute  for  each galaxy  the
redshift as  "seen" by the  observer, the $r$-band  apparent magnitude
and $M_{^{0.1}r}$,  the $r$-band absolute  magnitude of the  galaxy at
$z=0.1$.   Finally, we  mimic the  position-dependent  completeness by
randomly eliminating galaxies using the completeness masks provided in
the {\tt Sample dr4}.

We produce 10  mock catalogues from each SAM  catalogue, from which we
then select luminosity samples in the same way as the real sample.  As
pointed  out  by  \cite{Yang-04},  the  $L_{300}$  catalogue  is  only
complete  down  to  $M_{b_J}\approx -18.4$  (i.e.  $M_{^{0.1}r}\approx
-19.3$),   while  the   $L_{100}$  catalogue   is  complete   down  to
$M_{b_J}\approx -14$  (i.e. $M_{^{0.1}r}\approx -14.9$)  because it is
based  on higher-resolution  simulation.  This  implies that  the mock
samples constructed  from the $L_{300}$ catalogue  would be incomplete
out to  a distance of $\sim  350\mpch$.  To overcome  this problem, we
combined  the  $L_{100}$  and  $L_{300}$  mock  samples  by  selecting
galaxies  with  $M_{^{0.1}r}<-19$   from  the  $L_{100}$  samples  and
selecting those with $M_{^{0.1}r}>-19$ from the $L_{300}$ ones.

In total we  have 20 mock catalogues: 10  from the $L_{500}$ catalogue
and    10   from    the   $L_{100}$    plus    $L_{300}$   catalogues.
Figure~\ref{fig:ra_red} shows the  equatorial distribution of galaxies
in one  of the $L_{500}$  mock catalogues (middle)  and in one  of the
$L_{100}+L_{300}$  catalogues  (right), compared  to  the real  sample
(left).   The numbers of  galaxies in  our $L_{500}$  mock catalogues,
300,000  on average with  a dispersion  of $\sim7000$,  are consistent
with the observational sample.  In case of $L_{100}+L_{300}$, however,
the  numbers are  smaller: 250,000  on  average with  a dispersion  of
$\sim3700$.  As can be seen from Figure~\ref{fig:lf}, the model of K05
predicts fewer $L^\ast$ galaxies than the observations.

\section{Comparisons between models and observations}
\label{sec:comparison}

\begin{figure*}
\centerline{\epsfig{figure=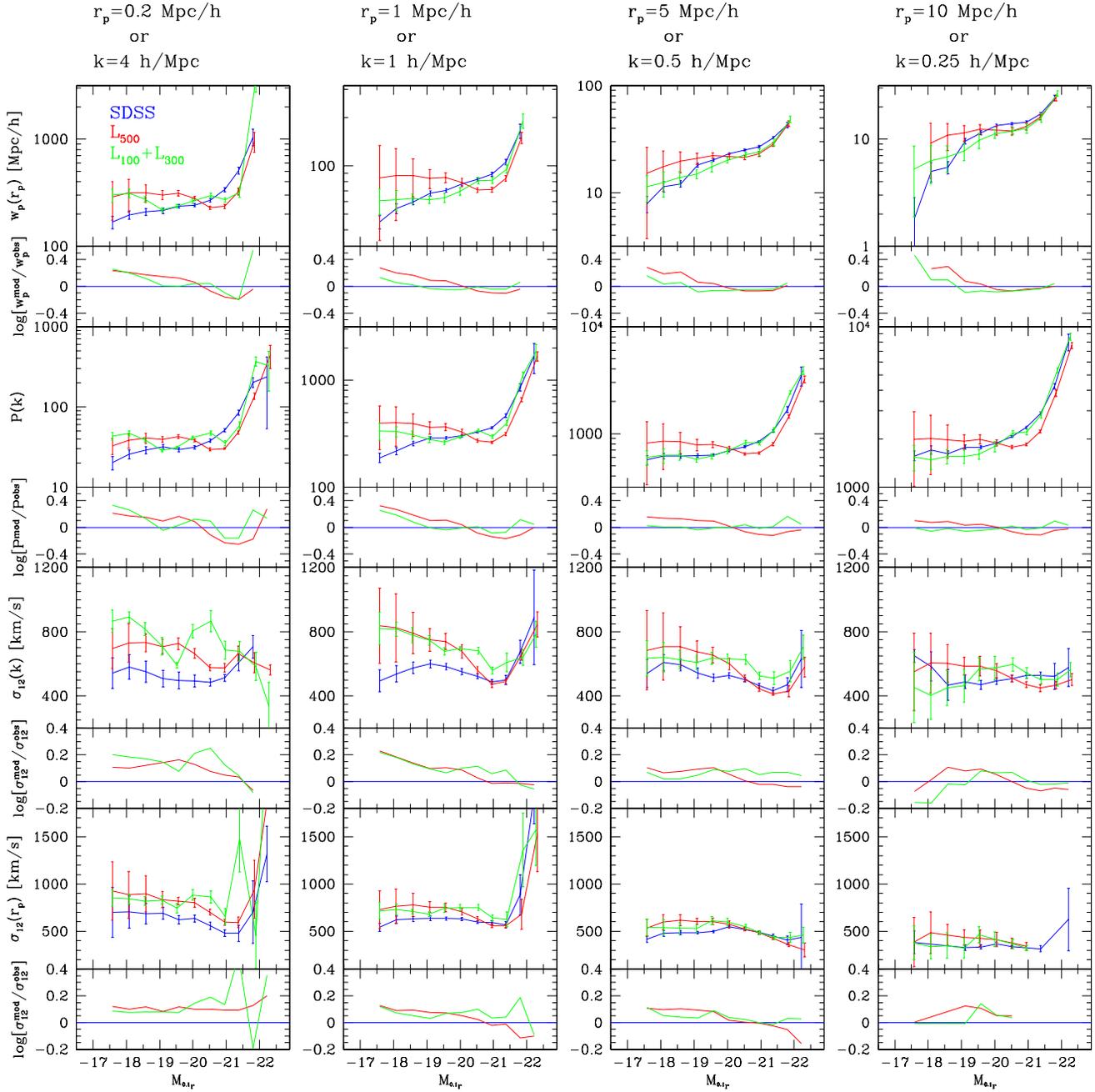,width=\hdsize}}
\caption{Clustering   and  velocity  statistics   as  a   function  of
luminosity on different scales, compared between model predictions and
observations.   Panels  from left  to  right  correspond to  different
projected separations $r_p$ or scales  $k$, as indicated above the top
panels. Panels from top  to bottom correspond to different statistics:
$w_p(r_p)$, $P(k)$, $\sigma_{12}(k)$  and $\sigma_{12}(r_p)$. Blue and
green lines are the  model predictions respectively from the $L_{500}$
and the  $L_{100}+L_{300}$ mock catalogues,  while blue lines  are for
the SDSS  DR4 observations.  The  smaller panel below each  bigger one
plots the ratios of the model prediction to the observation.  The PVDs
measured at $k=1\mpci$  are also compared to the  2dFGRS result (black
circles with error bars) presented by JB04.}
\label{fig:dr4_mock_scales}
\end{figure*}
\begin{figure*}
\centerline{\epsfig{figure=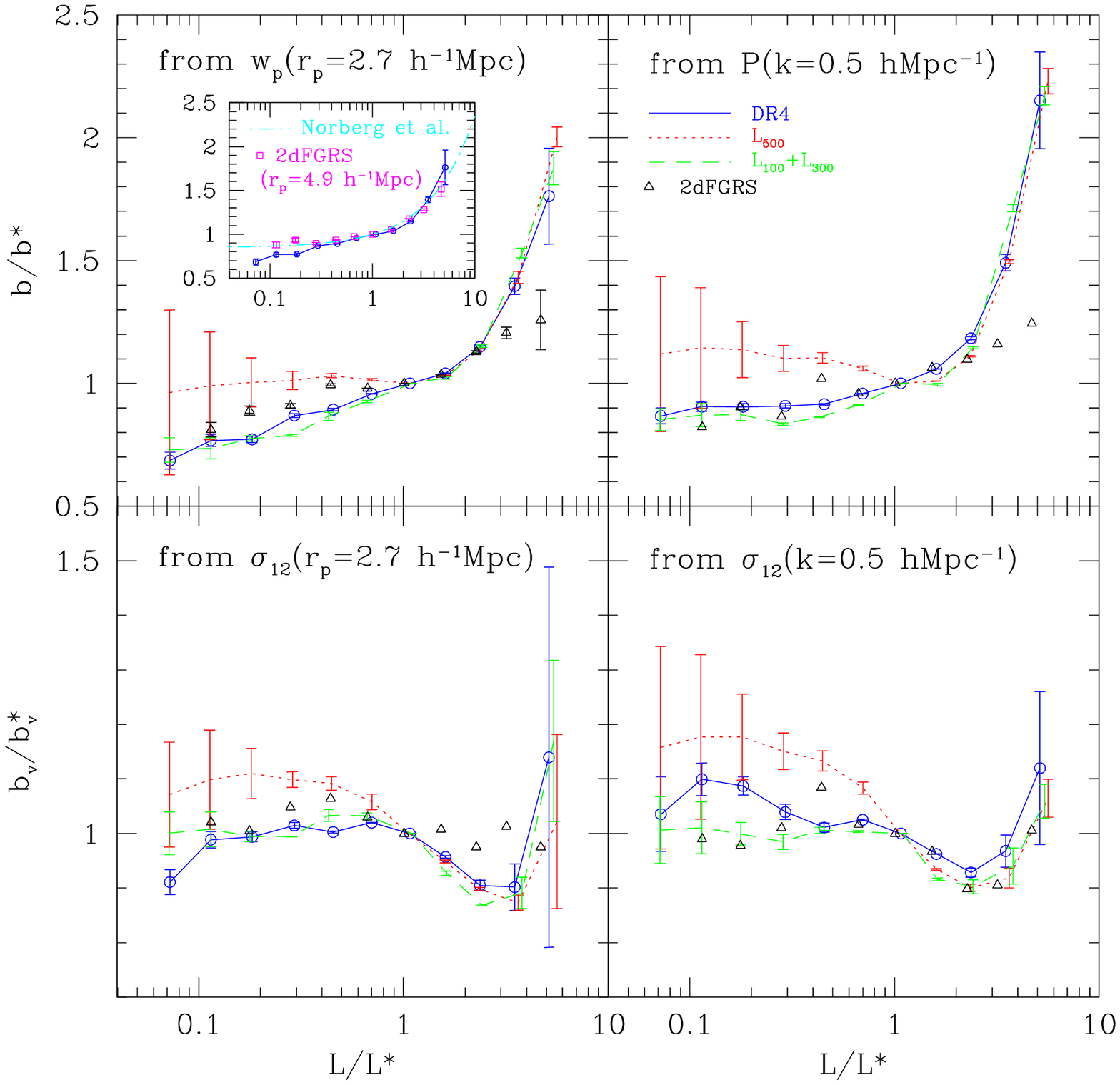,width=0.8\hdsize}}
\vspace{-0.5cm}
\caption{The spatial and velocity bias factors, scaled by their values
at  the characteristic  luminosity  $L^\ast$, as  a  function of  the
luminosity. The  bias factors are  estimated from the  measurements of
clustering/PVD at $r_p=2.7\mpch$ or $k=0.5\mpci$, as indicated in each
panel.  The  green data points  with error bars connected  with dashed
lines are  for the model  of Kang et  al., and those in  red connected
with dotted lines are for the model of Croton et al.. The results from
the  SDSS  are  represented  with  the blue  points  with  error  bars
connected  with the  solid lines.  The open  triangles are  the 2dFGRS
results from JB04. In the inset of the top panel at the left-hand, the
squares  with error bars  are also  from JB04  but are  obtained using
$w_p(r_p)$ at $r_p=4.89\mpch$ , and the dotted-dashed line is a fit to
$w_p(r_p)$   measurements    at   $r_p=4.89\mpch$   in    the   2dFGRS
$b/b^\ast=0.85+0.15L/L^\ast$ given in \citep{Norberg-01}. For clarity,
the  error bars  for the  2dFGRS results  (open triangles),  which are
comparable to that in the top-left panel, are not plotted in the other
panels.}
\label{fig:bias}
\end{figure*}

For each mock sample, we measure $w_p(r_p)$, $P(k)$ , $\sigma_{12}(k)$
and $\sigma_{12}(r_p)$ using the  same method as for the observational
samples   (\S~\ref{sec:obs}).   Figure~\ref{fig:dr4_mock}   shows  the
results   in    six   luminosity    intervals,   the   same    as   in
Figure~\ref{fig:dr2_dr4}.  In  each panel, the  average measurement is
plotted in  red for the  $L_{500}$ mock samples  and in green  for the
$L_{100}+L_{300}$  samples.  The error  bars indicate  the uncertainty
due to  cosmic variance as  estimated from 10 different  mock samples.
For  comparison,   the  observational  measurements   (blue  lines  in
Figure~\ref{fig:dr2_dr4})  are plotted  in this  figure, also  as blue
lines.  It should  be pointed out that, for  faint galaxies, the error
bars  on the  $L_{100}+L_{300}$ curves  are smaller  than that  on the
$L_{500}$ curves.  This is because  the faint galaxies are  taken from
the $L_{100}$ box, which artificially reduces the cosmic variance.

It is  seen that both  models match the observations  reasonably well.
The agreement  is better for  the two-point correlation  function than
for the  PVD, and it is  also better for more  luminous galaxies.  For
galaxies brighter  than -19, the  models reproduce the  $w_p(r_p)$ and
$P(k)$  measurements  on scales  of  $r_p>1\mpch$  or $k<1\mpci$,  but
marginally overpredict or underpredict the clustering power on smaller
scales  in some  cases.  For  galaxies fainter  than -19,  both models
predict   stronger  clustering   on   all  scales   compared  to   the
observations.  It  should be  noted  that  the  errors due  to  cosmic
variance are large so the disagreement is only marginally significant.

Similar result  are found for the  PVD.  On scales  of $r_p>1\mpch$ or
$k<1\mpci$  and above  $M_{^{0.1}r}=-20$, both  $\sigma_{12}(r_p)$ and
$\sigma{12}(k)$ are  well matched by the models  (especially the model
of C06). For $L^\ast$ galaxies (Sample L7 in Table~\ref{tbl:samples}),
the PVD values predicted by the  model of K05 are larger than those by
C06,  with   the  difference   becoming  more  significant   on  small
scales. This  can be understood because  K05 adopted in  their model a
larger  value  of  $\Omega_m$   than  C06.  For  faint  galaxies,  the
discrepancy between the models  and the observation seen in clustering
statistics is also  seen in the PVD. The  model predictions are higher
than the observations, but there are large uncertainties due to cosmic
variance.

These      results      are       shown      more      clearly      in
Figure~\ref{fig:dr4_mock_scales}, where we have plotted $w_p(r_p)$ and
$\sigma_{12}(r_p)$  at  $r_p=0.2,  1,  5,  10\mpch$,  and  $P(k)$  and
$\sigma_{12}(k)$  at  $k=0.25,  0.5,  1,  4\mpci$, as  a  function  of
absolute magnitude.  The ratios of  the model predictions  relative to
the SDSS observations  are also plotted We see  that both models match
the observations for high-luminosity galaxies ($M_{^{0.1}r}<-19$), but
overpredict the  clustering amplitude at low  luminosities.  The model
of K05 better reproduces the clustering statistics, while the model of
C06 more closely matches the  PVD.  Although the models predict higher
pairwise   velocities  at   faint  luminosities   than  seen   in  the
observations,  it is still  encouraging to  see that  the non-monotonic
dependence  of $\sigma_{12}(k)$  on  luminosity is  recovered by  both
models.   This behaviour also  exists in  configuration space,  but is
less  pronounced compared  to  Fourier space.   This is  qualitatively
consistent with the HOD results of \cite{Tinker-06}.

In the  two top panels of  Figure~\ref{fig:bias}, we plot  the bias of
galaxies $b$ as  a function of luminosity; this has  been done in many
previous papers (e.g. \citealt{Norberg-02,Tegmark-04,Zehavi-05}; Paper
I).  Here  the  bias  factor   is  normalised  by  its  value  at  the
characteristic  luminosity  $M_{\ast}$.  In  the top  left  panel,  we
estimate $b$ using the  projected 2PCF $w_p(r_p)$ at $r_p=2.7\mpch$ as
in  \cite{Zehavi-05}.  In  the  top  right panel,  we  use  $P(k)$  at
$k=0.5\mpci$ to  measure $b$.  The prediction of  K05 is  in excellent
agreement with  the observations. C06  predicts too strong a  bias for
faint galaxies.  In an analogous  way, we plot the velocity bias $b_v$
at  $r_p=2.7\mpc$  and   $\sigma_{12}(k)$  at  $k=0.5\mpci$  (  bottom
panels). This plot confirms that the non-monotonic behaviour found for
$\sigma_{12}(k)$   at   $k=1\mpci$  also   exists   at  other   scales
($k=0.5\mpci$)  and in  configuration space.  Again we  find  that the
model of K05  matches the observations very well,  while the C06 model
has  a steeper luminosity  dependence than  observed.  Note  that when
carrying out  these comparisons, we have normalised  the velocity bias
by the value $b_v^{\ast}$ at $M_{\ast}$. In fact, $\sigma_{12}$ in the
semi-analytic models  is $\sim 1.3$  (K05) and $\sim 1.1$  (C06) times
higher   than  in   the  observations   (Fig.\ref{fig:dr4_mock}).  One
possibility  is   that  a   a  CDM  model   with  a  lower   value  of
$\Omega_m^{0.6} \sigma_8$ would fits the observational results better.
However, as we will show in  the next section, this is not required by
the data.

Finally, it is interesting to compare measurements of the PVD from the
SDSS and  from the 2dFGRS,  as this will  indicate to what  extent the
observational  results   are  still  affected   by  variations  between
different  regions  of the  sky.   In  Figure~\ref{fig:bias} the  bias
factors  from  the 2dfGRS  calculated  by  JB04  are plotted  as  open
triangles.   From  the  figure,  we  see  that  there  are  small  but
significant differences with the  SDSS results.  For galaxies brighter
than $2L^\ast$, the spatial bias is  smaller in the 2dFGRS than in the
SDSS. This is surprising because it has been claimed in the literature
that there  is no significant  difference between the bias  factors in
the  two  surveys  (e.g.  \citealt{Zehavi-05};  Paper I).  We  note  ,
however, that Zehavi compared her SDSS results with the 2dFGRS results
of  \cite{Norberg-01}   ,  where  the   $w_p(r_p)$  measurements  were
normalised at $r_p=4.89\mpch$ and not at $2.7\mpch$. We have gone back
to  the 2dFGRS  data and  we have  estimated $b$  using  $w_p(r_p)$ at
$r_p=4.89\mpch$ and  we plot  the results in  Figure~\ref{fig:bias} as
squares  (the  inset of  the  top-left  panel).   For comparison,  the
fitting  function of  Norberg et  al.  is plotted  as a  dotted-dashed
line. As can  be seen, our results calculated at  $r_p =4.89 \mpch$ are
now perfectly consistent with Norberg et al, and are also in agreement
with the  SDSS results  at high luminosities.   This implies  that the
clustering properties of galaxies in  the two surveys have a different
dependence not  only on luminosity, but  also on scale.   We have also
studied the  results at  a variety of  different scales.  For example,
when $r_p=1\mpch$ or $k=1\mpci$ is used for estimating $b$ values, the
two  surveys are  perfectly consistent  with each  other  for galaxies
brighter than  $-19$, but for  fainter galaxies, both the  spatial and
the velocity biases are larger in the 2dFGRS than in the SDSS.

In  spite of  these complications,  we  find it  encouraging that  the
semi-analytic  models  can  reproduce  the qualitative  shape  of  the
luminosity dependence at magnitudes $M_{0.1r}<-19$.  It is not trivial
to achieve this success.  Previous generations of semi-analytic models
could  not  reproduce  the  strong  increase  of  clustering  at  high
luminosities \citep{Kauffmann-99,Norberg-01}.  For faint galaxies, the
clustering  statistics are  still not  entirely reliable,  because the
surveys are still being affected by cosmic variance. Larger samples or
better  estimation  methods  are  needed  in  order  to  make  further
progress.

\section{On the discrepancies at the faint end}
\label{sec:faint}

\begin{figure}
\vspace{-0.2cm}
\centerline{\epsfig{figure=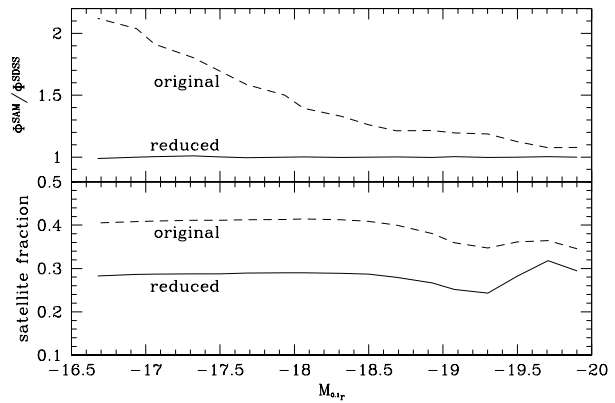,width=\hssize}}
\vspace{-2.5cm}
\caption{Top: the ratio of  the luminosity function from the $L_{500}$
model  catalogue relative  to that  from  the SDSS  DR4.  Bottom:  the
fraction of the  satellite population in the same  model catalogue. In
both  panels, the  dashed (solid)  line represents  the  result before
(after) reducing the satellite fraction (see the text for details).  }
\label{fig:frac2}
\end{figure}

\begin{figure*}
\vspace{-0.2cm}
\centerline{\epsfig{figure=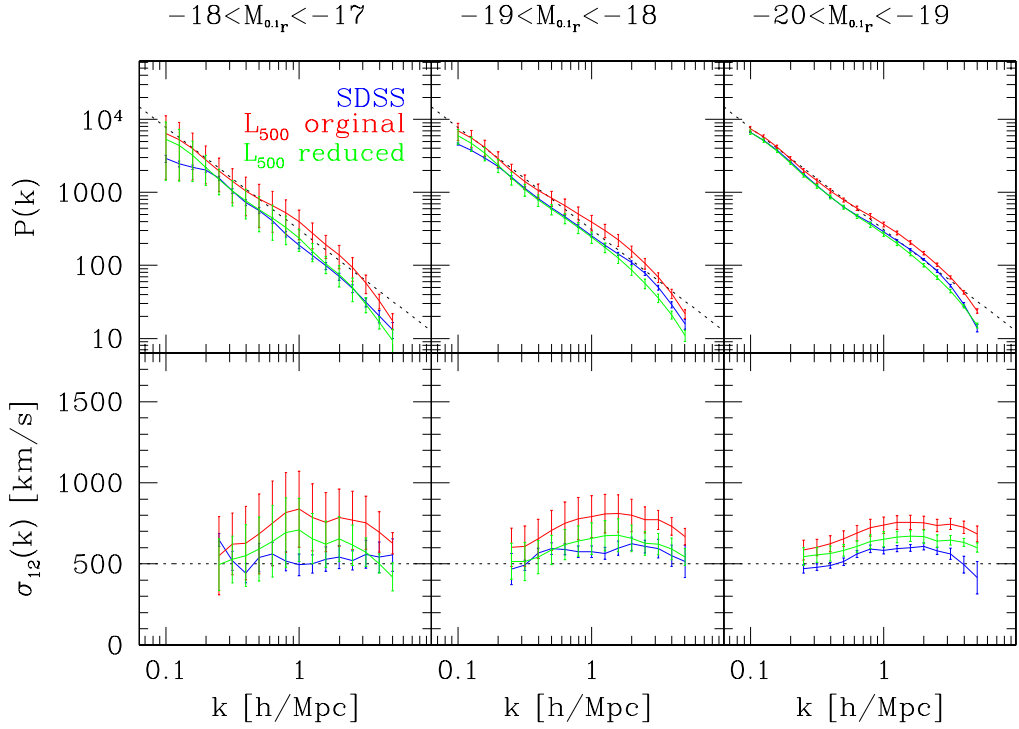,width=0.8\hdsize}}
\vspace{-4cm}
\caption{The power spectrum $P(k)$  (top panels) and the $k$-space PVD
$\sigma_{12}(k)$  (bottom  panels)  obtained  from  the  mock  samples
constructed based  on the $L_{500}$ model catalogue  without (red) and
with (green)  the satellite fraction  being reduced (see the  text for
details). The SDSS results are plotted in blue for comparison.}
\label{fig:dr4_mock_reduced}
\end{figure*}
\begin{figure*}
\centerline{\epsfig{figure=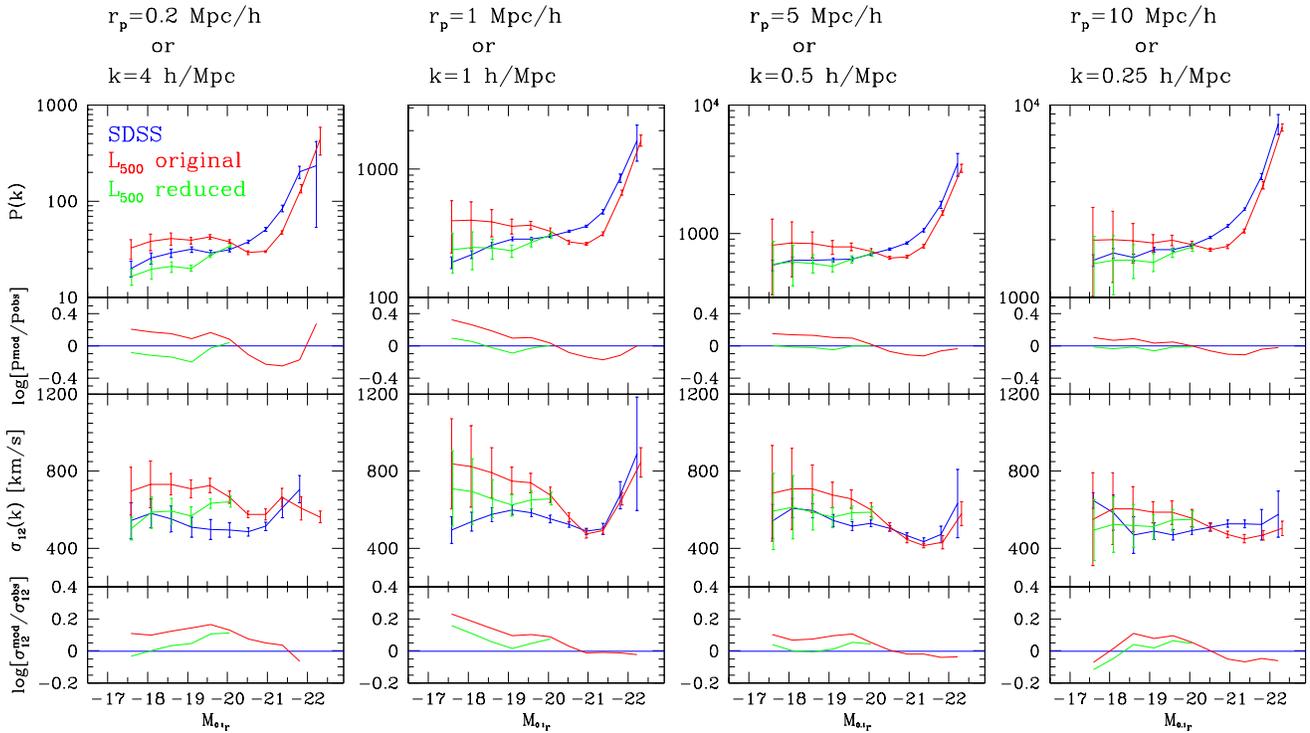,width=\hdsize}}
\vspace{-7.7cm}
\caption{$P(k)$ and  $\sigma_{12}(K)$ as  a function of  luminosity on
different scales, compared between  the mock samples constructed based
on the  $L_{500}$ model catalogue  without (red) and with  (green) the
satellite fraction being reduced (see  the text for details). The SDSS
results are plotted  in blue for comparison.  The  smaller panel below
each  bigger one  plots  the ratios  of  the model  prediction to  the
observation.}
\label{fig:dr4_mock_reduced_scales}
\end{figure*}

\begin{figure*}
\centerline{\epsfig{figure=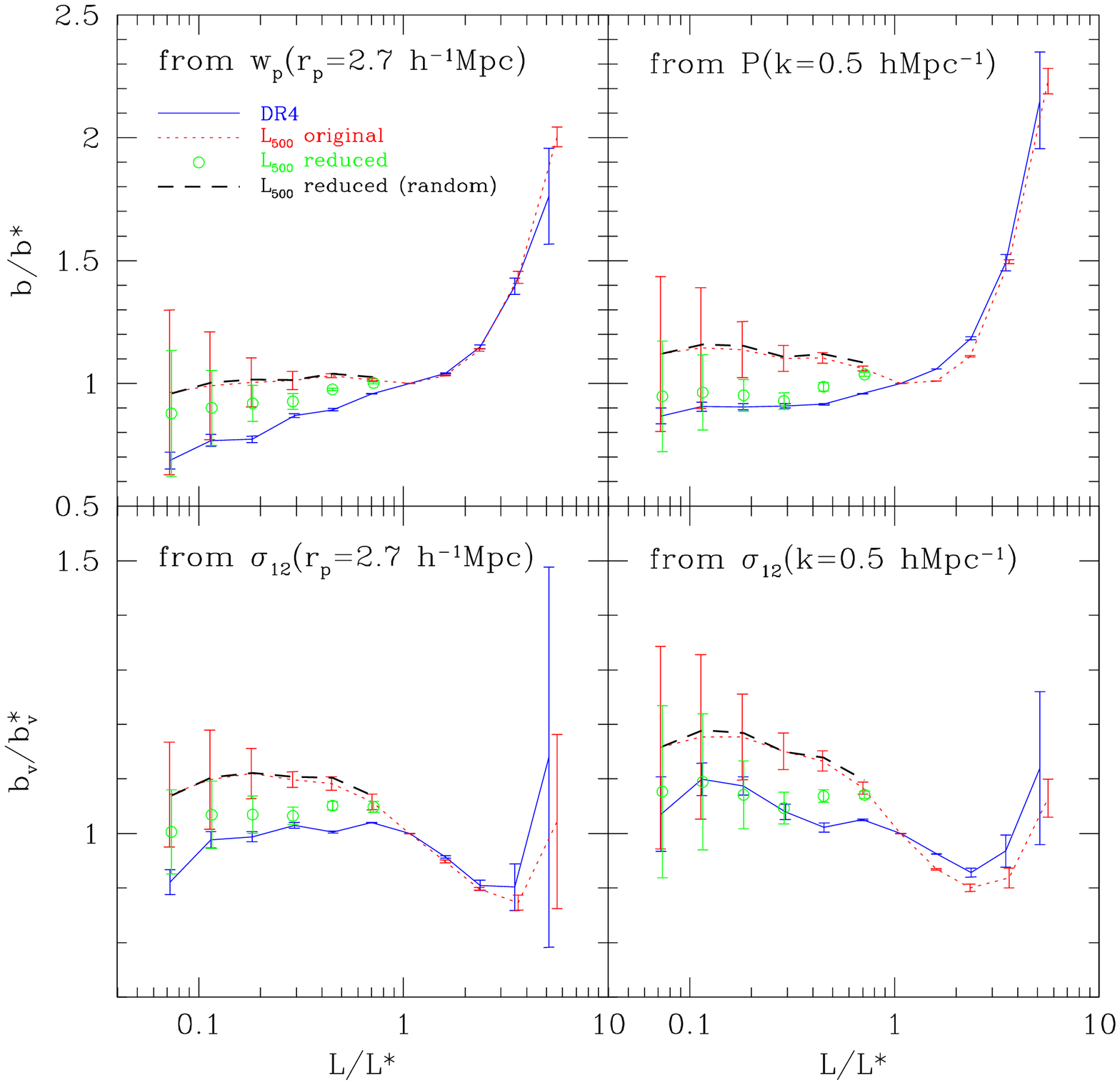,width=0.8\hdsize}}
\vspace{-0.5cm}
\caption{The spatial  and velocity relative bias factor  as a function
of the luminosity, compared between the mock samples constructed based
on the  $L_{500}$ model catalogue  without (red) and with  (green) the
satellite fraction being reduced (see  the text for details). The SDSS
results are plotted  in blue for comparison. The  black dashed line is
for the mock samples in which the total number of galaxies was reduced
(to  match the luminosity  function) but  the satellite  fraction kept
unchanged.}
\label{fig:bias_reduced}
\end{figure*}

As we have seen  in \S\ref{sec:comparison}, there are some significant
discrepancies  between the  observed  PVD of  faint  galaxies and  the
result   of   the   semi-analytic   models.    We   have   also   seen
(Figure~\ref{fig:lf}) that both the C06 and K05 models overpredict the
number  of  galaxies at  the  faint  end  of the  luminosity  function
($M_{^{0.1}r}>-20$).   This   is  the  luminosity   regime  where  the
disagreement with  the PVD data is  worst.  It is  thus interesting to
ask whether reducing the number  of faint galaxies to provide a better
match to the  luminosity function would, at the  same time, also solve
the PVD discrepancy.

To  answer  this  question,  we  have performed  several  simple  mock
experiments.  In the first experiment,  we randomly remove a number of
faint  galaxies  with   $M_{^{0.1}r}>-20$  from  the  $L_{500}$  model
catalogue so that the  resulting catalogue has the same $^{0.1}r$-band
luminosity   function   as   the   SDSS  observations   presented   in
\citet{Blanton-03-LF}.  When computing the luminosity function for the
model  catalogue, we  have corrected  the $r$-band  absolute magnitude
$M_{r}$ of each model galaxy to its $z=0.1$ value $M_{^{0.1}r}$ in the
same way as  described in \S3.  We construct  10 mock catalogues using
using this reduced catalogue and  we analyse the clustering and PVD in
the same way as in \S4.  We  find that the results are almost the same
as presented in \S4.

Since the  PVD reflects the  action of the local  gravitational field,
the discrepancies  in the PVD at  the faint end imply  that the models
predict  too  many  faint  galaxies  that are  located  in  high  mass
haloes. As we will  see (Figure~\ref{fig:mvir_hist}), these are mainly
{\it satellite} systems rather than  the central galaxies of their own
halo.  It  is thus  natural to speculate  that it is  these satellites
that  are   responsible  for  the   very  large  PVD  values   at  low
luminosities.  We thus  repeat  the above  experiment  except that  we
preferentially   eliminate   satellites.    Satellite  galaxies   with
$M_{^{0.1}r}>-20$ are  randomly removed until  the luminosity function
comes into  agreement with the  observation, or until the  fraction of
satellite galaxies is reduced by  more than 30\%.  In the latter case,
we further remove  a number of central galaxies at  random so that the
resulting  catalogue   has  the   same  luminosity  function   as  the
observation.

Figure~\ref{fig:frac2}  compares  the   luminosity  function  and  the
satellite fraction for the  original and the reduced model catalogues.
Figure~\ref{fig:dr4_mock_reduced}  compares the power  spectrum $P(k)$
and the Fourier  space PVD $\sigma_{12}(k)$ for the  original (red) and
the reduced (green)  catalogues. The SDSS results are  plotted in blue
for comparison.  Figure~\ref{fig:dr4_mock_reduced_scales} plots $P(k)$
and $\sigma_{12}(k)$  at $k=0.25, 0.5, 1,  4 \mpci$, as  a function of
absolute  magnitude.  In  Figure~\ref{fig:bias_reduced},  we plot  the
results for  the spatial  and velocity bias  factors.  As can  be seen
from  the three figures,  both the  clustering power  and the  PVD for
faint galaxies  are reduced substantially and change  to be consistent
with   the  SDSS   results.    For  comparison,   we   also  plot   in
Figure~\ref{fig:bias_reduced}  (dashed black  lines) the  bias factors
obtained  from the  first  experiment  in which  the  number of  faint
galaxies  is  reduced  at  random,  independent of  whether  it  is  a
satellite or  a central galaxy.  As can  be seen , there  is almost no
effect on the results.  Finally we have also investigated what happens
if we allow the fraction of  satellite galaxies to be reduced by up to
50\%. The agreement with observations is no longer very good; both the
clustering amplitude and the peculiar velocities are now too small.

\section{On the non-monotonic luminosity dependence  of the PVD}
\label{sec:bimodal}

\begin{figure*}
\centerline{\epsfig{figure=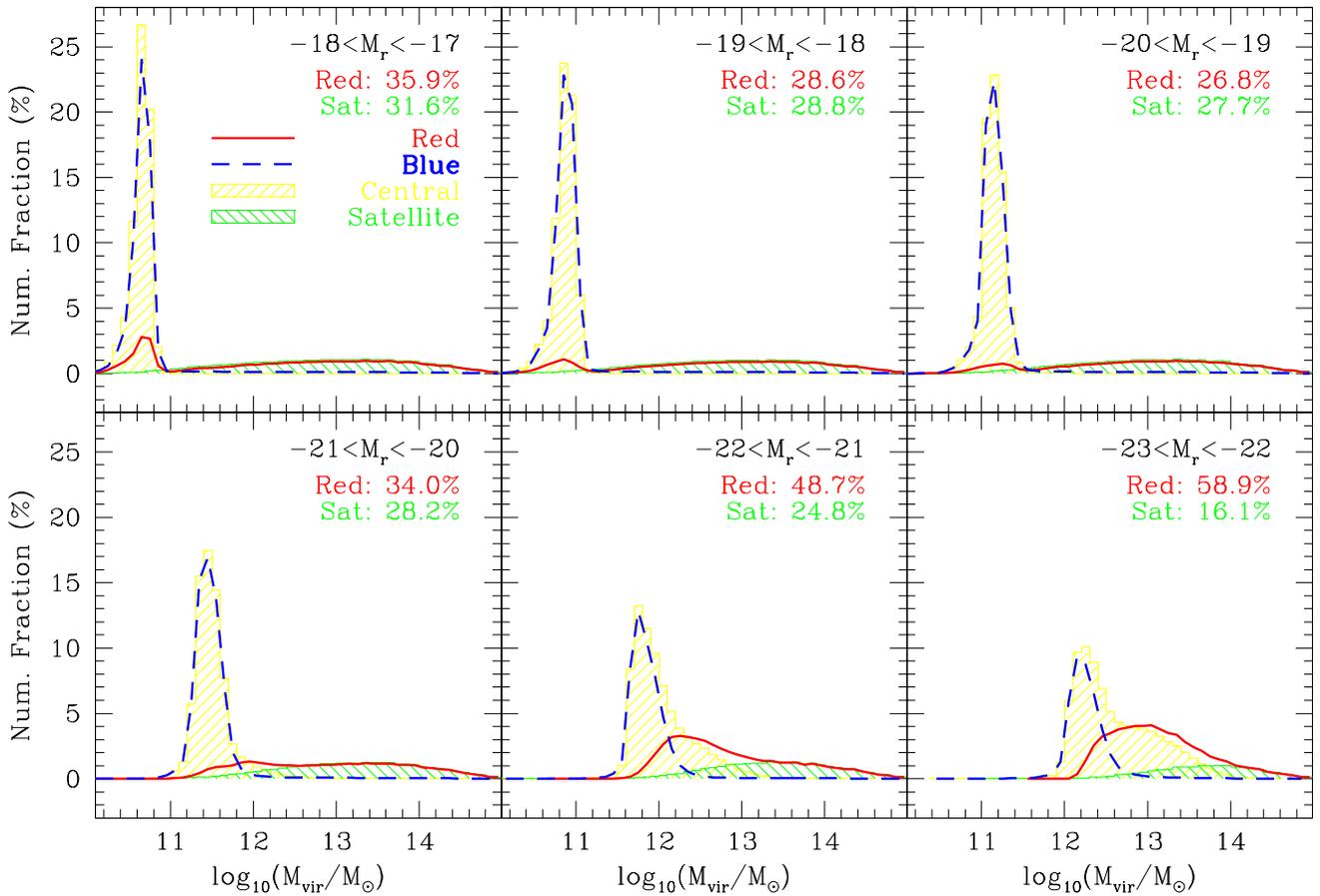,width=\hdsize}}
\vspace{-5cm}
\caption{Distribution of  the virial mass  of host dark  matter haloes
for  model galaxies in  the reduced  $L_{500}$ catalogue  in different
luminosity intervals, as indicated in each panel.
The  red solid  (blue dashed)  lines represent  the red  (blue) galaxy
population, and  the yellow (green) shaded histogram  shows the result
for central  (satellite) galaxies.  The fraction of  these populations
are indicated in each panel.}
\label{fig:mvir_hist}
\end{figure*}

The  non-monotonic  luminosity  dependence  of PVD  indicates  that  a
substantial fraction  of faint galaxies must reside  in high-mass dark
matter haloes.  In paper II,  we discussed how the faint red satellite
galaxy population in dense  environments, even though small in number,
can still dominate  the PVD on small scales  ($k\sim1\mpci$).  Here we
use  the galaxy  catalogues  from the  semi-analytic  models to  check
whether this hypothesis is correct.

We  first divide  the model  galaxies in  the {\em  reduced} $L_{500}$
catalogue  into  different  luminosity  intervals (we  use  rest-frame
magnitudes for this analysis).   We then divide each luminosity sample
into red and blue  subsamples using a luminosity-dependent colour cut,
which is  determined using the colour-magnitude diagram  of galaxies in
the $L_{500}$  catalogue. The colour  distribution is bimodal,  so the
natural  place   to  divide  the   galaxies  into  "red"   and  "blue"
subpopulations is at  the minimum between the two  peaks in the colour
distribution.

Figure~\ref{fig:mvir_hist} shows  the distribution of  the virial mass
of the  host dark matter  haloes for galaxies in  different luminosity
intervals.   Red solid  and blue  dashed lines  are for  red  and blue
galaxies.  The fraction of red and satellite populations are indicated
in each panel.  We also plot the the results for central and satellite
galaxies.

In each  luminosity interval, the virial  mass of host  haloes shows a
peak at lower masses and a longer tail to higher masses.  The position
of the first  peak moves to higher masses  for more luminous galaxies.
When the galaxies  are divided into central and  satellite systems, we
see that the central galaxies dominate the first peak at low halo mass
and  the satellite galaxies  are located  in the  tail of  higher mass
halos.   This result may  provide clues  to understanding  the bimodal
colour distribution of galaxies.  The fraction of satellite systems in
high mass halos increases with decreasing galaxy luminosity up to $M_r
\sim -19$, and then remains constant at around $\sim$ 30 \% at fainter
magnitudes.  It is this satellite population that gives rise to a high
PVD at the faint luminosities. In the models the satellites are mainly
red and we  note that the satellite fractions  predicted by the models
are in good agreement of the  fraction of red galaxies observed in the
SDSS (see Table 1 of Paper I).

\section{Summary}
In this paper, we have compared the clustering and pairwise velocities
for  galaxies in  different luminosity  intervals measured  from Sloan
Digital Sky Survey with results from mock catalogues constructed using
the semi-analytical  models of \citet{Kang-05}  and \citet{Croton-06}.
We show  that the  models can match  a number  of key features  of the
luminosity  dependence of the  clustering and  the PVD,  including the
monotonic increase of the clustering amplitude with luminosity and the
non-monotonic behaviour of the PVD.  PVD.

A direct look into the  galaxy catalogues supports the conclusion that
a substantial fraction of faint galaxies must reside in high mass dark
matter  haloes.   The  luminosity  dependence  of the  PVD  is  mostly
determined by  how galaxies of different  luminosities are distributed
among/inside  dark matter  haloes.  All  these results  are consistent
with   the   recent  studies   of  \cite{Slosar-Seljak-Tasitsiomi-06},
\cite{Tinker-06} and \cite{vandenBosch-06}  which   were  carried  out
using halo occupation distribution (HOD) models. 

We  have also  identified a  few significant  differences  between the
models  and the  observations. The  differences are  generally  at the
level of a few tens of  percent both in the clustering and in velocity
statistics. One difference is that  the PVD predicted by the models is
systematically  higher than the  observations.  Another  difference is
that the clustering of faint galaxies, especially in the C06 model, is
significantly  stronger than that  observed in  the SDSS.  However, we
note  that cosmic  variance  effects are  still  significant at  faint
luminosities    because    the    effective   surveyed    volume    is
small. Significant differences also still exist between the 2dFGRS and
the  SDSS  clustering measurements.   If  this  overprediction of  the
clustering  for faint  galaxies is  confirmed, our  experiment  in \S5
shows that the  fraction of faint satellite galaxies  in massive halos
will have  to be reduced by a  factor of $\sim30\%$ in  order to bring
the models  into better agreement with  the data. The  recent study by
\citet{Weinmann-06},  which  compares  the  fraction  of  central  and
satellite galaxies in  dark halos between the C06  model and the SDSS,
has  found that  the fraction  of the  faint galaxies  is too  high in
massive halos. The strong clustering  found here for faint galaxies in
the model is clearly consistent with their findings.

\section*{Acknowledgments}

CL acknowledges the financial  support by the exchange program between
Chinese Academy of Sciences and  the Max Planck Society.  This work is
supported  by  NSFC(10643005,  10373012,  10533030), by  Shanghai  Key
Projects  in Basic  research (04jc14079,  05xd14019), and  by  the Max
Planck Society.

The Millennium  Run simulation used in  this paper was  carried out by
the  Virgo Supercomputing Consortium  at the  Computing Centre  of the
Max-Planck Society in Garching.  The semi-analytic galaxy catalogue is
publicly available at http://www.mpa-garching.mpg.de/galform/agnpaper.

Funding for  the SDSS and SDSS-II  has been provided by  the Alfred P.
Sloan Foundation, the Participating Institutions, the National Science
Foundation, the  U.S.  Department of Energy,  the National Aeronautics
and Space Administration, the  Japanese Monbukagakusho, the Max Planck
Society, and  the Higher Education  Funding Council for  England.  The
SDSS Web  Site is  http://www.sdss.org/.  The SDSS  is managed  by the
Astrophysical    Research    Consortium    for    the    Participating
Institutions. The  Participating Institutions are  the American Museum
of  Natural History,  Astrophysical Institute  Potsdam,  University of
Basel,   Cambridge  University,   Case  Western   Reserve  University,
University of Chicago, Drexel  University, Fermilab, the Institute for
Advanced   Study,  the  Japan   Participation  Group,   Johns  Hopkins
University, the  Joint Institute  for Nuclear Astrophysics,  the Kavli
Institute  for   Particle  Astrophysics  and   Cosmology,  the  Korean
Scientist Group, the Chinese  Academy of Sciences (LAMOST), Los Alamos
National  Laboratory, the  Max-Planck-Institute for  Astronomy (MPIA),
the  Max-Planck-Institute  for Astrophysics  (MPA),  New Mexico  State
University,   Ohio  State   University,   University  of   Pittsburgh,
University  of  Portsmouth, Princeton  University,  the United  States
Naval Observatory, and the University of Washington.


\bsp
\label{lastpage}

\end{document}

%% file: macro1.tex
\newcommand{\etal}{{et al.~}}

\newcommand{\absmag}{M_{^{0.1}r} - 5 \log_{10} h}

\newcommand{\kms}{\>{\rm km}\,{\rm s}^{-1}}

\newcommand{\mpc}{\>{\rm Mpc}}

\newcommand{\msun}{\>{\rm M_{\odot}}}

\newcommand{\beq}{\begin{equation}}
\newcommand{\eeq}{\end{equation}}

\newcommand{\mpch}{\>h^{-1}{\rm {Mpc}}}
\newcommand{\mpci}{\>h{\rm {Mpc}}^{-1}}

\newcommand{\msunh}{\>h^{-1}\rm M_\odot}

\newcommand{\apj}{ApJ}
\newcommand{\apjs}{ApJS}

\newcommand{\aj}{AJ}
\newcommand{\mnras}{MNRAS}

\newcommand{\nat}{Nature}
\newcommand{\physrep}{Phy. Rep.}

\newdimen\hssize
\newdimen\hdsize 